# Hierarchical Aggregation Clustering Algorithms Derived from the Bi-partial Objective Function

Jan W. Owsiński
Systems Research Institute, Polish Academy of Sciences
owsinski@ibspan.waw.pl

Abstract

The paper outlines the principles of construction of a broad class of hierarchical aggregation algorithms of cluster analysis, essentially based on minimum distance mergers, which are derived from the general bi-partial objective function. It is shown how the algorithms arise from the bi-partial objective function, their affinity with the classical hierarchical aggregation algorithms is demonstrated, and the examples of such algorithms for the concrete forms of the bi-partial objective function are provided. This amounts to the first explicit and, at the same time, quite general, connection between optimization in clustering and the hierarchical aggregation algorithms. Thereby, the respective hierarchical algorithms gain a deeper justification, the means for evaluating the quality of clustering is provided, along with the criterion of stopping the cluster mergers.

## 1. Introduction: the clustering problem

Clustering of objects or observations is the subject of a domain in the field of data analysis – cluster analysis – consisting in solving the problem of *dividing a set of observations or objects into subsets (clusters) in such a way that the items, belonging to the same cluster, are possibly similar or close to each other, while the items, belonging to different clusters, are possibly dissimilar, or distant*. This problem, apparently and intuitively quite obvious in its nature, is the basis of a very broad spectrum of human endeavors and, indeed, achievements. These range from language formation (meanings of words vs. objects, features and phenomena in reality) down to automatic self-control of autonomous vehicles (clusters of communication and control nodes). Yet, even though the problem forms a kind of intellectual paradigm, its concrete and precise formulation for the purpose of solving it in a practical situation, with definite data, is still missing.

There are, of course, dozens of clustering algorithms currently available, both in literature and in publicly available software libraries, based on a variety of presumptions and featuring highly diversified properties in terms of computational effort and the character of results. During the last half century a lot of books have been published, devoted to cluster analysis, some selected examples are Hennig et al. (2015), Everitt et al. (2011), Xu and Wunsch (2009) or Mirkin (1995). Many of these books have been published in several consecutive editions. These books present not only various algorithms, but also theoretical foundations for and analyses of the essential elements and questions, arising in formulation and solving of the clustering problem, as defined above.

Even a superficial consideration of this verbally expressed problem leads to the conclusion that these essential elements and questions are related to: (i) the measure(s) of distance and/or proximity, both for individual objects / observations, and (ii) for their sets (clusters); (iii) the formulation of an overall objective function, representing the quality of a given partition, in accordance with the content of the problem; and (iv) the corresponding algorithm that would lead to solution, or at least its approximation, for the objective function formulated.

The existing methodologies do not respond satisfactorily to these challenges, especially regarding points (iii) and (iv). This statement holds even if there exist algorithms that are very computationally effective and yield results which can be accepted on the basis of some "external" criteria, not related to these algorithms and expressing various kinds of rational qualifications of the partitions obtained.

In this paper we deal with one of the families of clustering algorithms, the hierarchical merger or aggregation algorithms. It is one of the most popular and deeply rooted groups of algorithms, of an obvious intuitive character. We shall show how algorithms of this kind can be derived from the objective function, called bi-partial (see Owsiński, 2020), which reflects adequately the essence of the clustering problem. This objective function is formulated in a very general manner, and then it is shown how for its more concrete forms the corresponding hierarchical merger algorithms can be derived from it, leading to its suboptimisation.

In this way, not only particular partitions can be evaluated in terms of the original clustering problem, but also simple algorithms can be designed, corresponding to the manner of evaluating these partitions, leading to suboptimal solutions. On the top of this, the stop condition for the algorithms is specified, and a room is established for the improvement of the suboptimal solutions.

In the next section, the hierarchical aggregation algorithms are characterized. Then, in Section 3, comments are forwarded on the relation between the functioning of the known algorithms from this group and the issue of clustering optimization. In Section 4 a telling example is provided for both the construction of the bi-partial objective function and the reasoning behind the derivation of the respective hierarchical merger algorithms. Section 5 provides the general formulation of the bi-partial objective function and the principles of design of the corresponding hierarchical merger algorithms. The next section, Section 6 is devoted to presentation of examples of concrete forms and the objective function and corresponding algorithms, along with some comments, mainly of technical character. Section 7, containing conclusions, closes the paper.

2. The hierarchical aggregation algorithms

Hierarchical aggregation algorithms belong among the most popular, classical clustering algorithms. They form a relatively broad class of algorithmic procedures, including single linkage (nearest neighbor), complete linkage, average linkage, Ward's algorithm, and yet quite a number of other ones, including their technical variants. All these algorithms work in such a way that for a set $X$ of $n$ objects or observations, indexed $i$ (or $j$), that is: $i \in \{1,…,n\} = I$, and somehow defined distance $d(.,.)$ for any two objects from the set $X$ (or the respective space, $E_X$, of which $X$ is a subset), one realizes the following generic procedure:

    (0) treat all the objects, indexed $1,…,n$ as $n$ separate clusters;

    (1) find the smallest of the distances between the pairs of clusters;

    (2) merge the corresponding two clusters;

    (3) check, whether the new number of clusters is still bigger than 1, if not – stop;

    (4) update the distances between clusters (with respect to the newly formed cluster);

    (5) go to step 1.

The particular algorithms from this group differ by the step 4 (distance matrix updating)[1], and it is exactly the essential source for the algorithmic variety. Lance and Williams (1966, 1967) were the first to systematize these algorithms through the parametric formula of inter-cluster distance updating. Later on, this formula was being yet extended in order to accommodate a broader variety of algorithms.

If we denote by $q$, $q = 1,…,p$ the index of clusters $A_q$ in a partition $P$, and by $D(A_q,A_{q'})$, or $D_{qq'}$, the distance between clusters, then the formula, proposed by Lance and Williams for inter-cluster distance updating, assuming the merged clusters are indexed $q^*$ and $q^{**}$, is as follows:

---

[1] Some of the variants may differ at other steps, e.g. step 1, where additional conditions besides minimum distance may be added.

$$D_{q^*\cup q^{**},q} = a_1 D_{q^*q} + a_2 D_{q^{**}q} + b D_{q^*q^{**}} + c|D_{q^*q} - D_{q^{**}q}| \qquad (1)$$

defining the distance between the new cluster $A_{q^*}\cup A_{q^{**}}$ and any other cluster $A_q$. Concrete values of coefficients $a_1$, $a_2$, $b$ and $c$ correspond to particular algorithms, as this is exemplified in Table 1 for some of the most popular ones. (Concerning this table, and, in general, $n_q$ denotes the number of objects in cluster indexed $q$.)

Over time, both new algorithms have been found to follow the Lance and Williams formula, with different values of the coefficients, and the formula itself has been extended to encompass more coefficients and hence also a much broader class of algorithms, e.g. by M. Jambu. For an interesting exposition a Reader is referred to Podani (1989) (another worthwhile reference is Murtagh and Contreras, 2012). The number of the thus described aggregation algorithms is now at about 20.

Table 1. Examples of hierarchical merger algorithms and their Lance-and-Williams formula coefficients

| Algorithm | $a_1$ | $a_2$ | $b$ | $c$ |
|---|---|---|---|---|
| Single linkage (nearest neighbor) | 1/2 | 1/2 | 0 | -1/2 |
| Complete linkage (farthest neighbor) | 1/2 | 1/2 | 0 | 1/2 |
| Unweighted average (UPGMA) | $n_{q^*}/(n_q+n_{q^*})$ | $n_{q^{**}}/(n_q+n_{q^{**}})$ | 0 | 0 |
| Weighted average (WPGMA) | 1/2 | 1/2 | 0 | 0 |
| Centroid (UPGMC) | $n_{q^*}/(n_q+n_{q^*})$ | $n_{q^{**}}/(n_q+n_{q^{**}})$ | $-n_{q^*}n_{q^{**}}/(n_{q^*}+n_{q^{**}})$ | 0 |
| Median (WPGMC) | 1/2 | 1/2 | -1/4 | 0 |

The generic procedure gives rise, in a natural manner, to a graph of consecutive mergers, a dendrogram (a tree), containing $n$ leaves, $n$-1 nodes (corresponding to mergers) and 2($n$-1) edges. The dendrogram is a very powerful representation of the concrete way of proceeding of the algorithm (and – hopefully – also of the dataset, whose essential characteristics ought to be reflected in this tree). It shows the groups of objects at various "levels of distinction" (length of distances between the clusters consecutively merged) and their interconnections, as well as a measure of cluster affinity in the form of the "height" of nodes, either ordinal (sequence of mergers) or given through the value of distance, at which the merger occurred.

In connection with the above, it must be emphasised that the hierarchical aggregation algorithms are very appealing as highly intuitive and easily interpretable. Yet, in view of the computational requirements of these algorithms, primarily the necessity of processing the matrices of distances, they are not appropriate for the large data sets, and so, if used for such data sets, it is usually in hybrid techniques, in which they are used in conjunction with some other methods, e.g. based on density of objects in space $E_X$.

Yet, there is another issue in the applicability of the hierarchical merger algorithms. Namely, although widely used in clustering, they actually do not solve the fundamental problem of cluster analysis, i.e., slightly reformulating the definition, given at the outset:

"to partition the set of objects $I$ into subsets (clusters) $A_q$, $q = 1,…$, so that the objects in the same clusters be possibly close (similar) to each other, while objects in different clusters – be possibly distant (dissimilar)".

In other words – these algorithms do not produce the partition, unless we apply an "external" criterion or method to, say, cut the dendrogram at a given height. And this is actually what is usually being done.

The present paper is meant to attempt bridging exactly this gap, namely – to show a class of objective functions for clustering, for which there exists (from which can be derived) a class of the hierarchical merger algorithms, which solve, at least in an approximate manner, the clustering problem as represented by these objective functions. This paper does not aim at mathematical precision, but is intended to present the essential principles and the logic of the approach.

The paper follows some of the considerations, presented in Owsiński (2020), concentrating, though, on the subject of relation between the objective functions and the hierarchical merger algorithms, and presenting some important new developments.

3. Hierarchical aggregation algorithms and optimality

Almost starting with the initial work of Florek et al. (1956), which introduced the algorithms in question (actually: the single linkage algorithm), the relation between these algorithms and some sort of optimality has been discussed. Particularly known has become the link between the single linkage algorithm and the minimum spanning trees (the Kruskal's algorithm). Yet, there is no comprehensive theory, nor even empirical study, showing the links of the entire class of the hierarchical merger algorithms, or its sub-classes, with definite optimality concepts.

When considering the relation between the algorithms in question and optimality, one can start with the very basic "*minimum distance*" heuristic of these algorithms. One deals here with an explicit *step-wise rationality*, oriented at some sort of quasi-optimisation, composed of two elements: (1) distance minimization (rather than any other choice of clusters for the merger), and (2) the distance updating rationality (the distance-related merger rule), expressed through the Lance-Williams-type formula. Notwithstanding various studies, showing the associations between the particular progressive merger algorithms and definite optimality concepts and objective functions (see further on) the entire class of procedures can be seen as representing various kinds of greedy-type "local optimality intuitions", having, in general, no properties, related to any broader optimality.

This changes, when we consider the *entire dendrograms*, rather than consecutive mergers. It then turns out that dendrograms, produced by various aggregation procedures, may be shown to have definite optimality properties. The relation between single linkage and minimum spanning tree generation algorithms was already mentioned. Similarly early (see Hartigan, 1967) the question was asked of the possibly best approximation of the actual distances between objects by the dendrograms, produced with different aggregation procedures (approximation of distance by the ultrametrics, corresponding more directly to dendrograms). In a similar vein, the concept of parsimony also appeared in the context of the dendrograms, although understood somewhat differently. This, however, concerns not the optimality of partitions (clustering results), but the possibly most accurate rendition of the distance structure in the data set by the dendrogram.

More recently, an approach was elaborated of a much broader meaning, based on the seminal paper by Dasgupta (2016), later on developed significantly by Cohen-Addad and associates (see, e.g., Cohen-Addad et al., 2017), to link hierarchical clustering with a definite class of objective functions at the level of entire dendrograms. This approach applies for a broad class of algorithms, on the one hand, and of objective functions, on the other, with appropriate conditions of correspondence being formulated.

We are, however, interested in the relation between the "*minimum distance merger*" hierarchical clustering algorithms and the objective function, or functions, pertaining to the *proper problem of clustering*, as formulated at the outset, i.e. of *obtaining an optimum partition*, and not a tree (i.e. dendrogram). Let us note, at this point, that there do exist, indeed, very many criteria or indices (dozens of them!) of clustering quality, see, e.g., Rendón et al. (2011), Vendramin, Campello and Hruschka (2010), Owsiński (2020), Section 5.4, or the choice, offered by the R environment (see

Cluster Analysis in R) documents[2]. None of these, though, is directly associated (being derived from put apart) to any of the hierarchical clustering algorithms, and they are used post-hoc to select the "best" clustering results among many that can be produced with various algorithms and their diverse parameterisations, with no relation whatsoever to the logic of the algorithm, whose results are being evaluated.

At this point we might quote from the already mentioned Dasgupta (2016): "These [hierarchical clustering algorithms] are widely used and are part of standard packages for data analysis. Despite this, there remains an aura of mystery about the kinds of clusters that they find. In part, this is because **they are specified procedurally rather than in terms of the objective functions they are trying to optimize**. For many hierarchical clustering algorithms, it is hard to imagine what the objective function might be." (emphasis added). Dasgupta (2016) tried to make the link for the entire dendrograms, while here we shall concentrate on the partitions, i.e. on the proper problem of clustering.

Yet, even with respect to partitions, there have been – besides the work by the present author, conducted since the early 1980s, and summarized in Owsiński (2020) – some early hints as to the link with hierarchical aggregation algorithms. The notable case is the one of Ducimetière (1970), where the connection between the average link algorithm and the objective function of clustering, proposed, within a broader framework, in Rubin (1967), is indicated. Yet, this very important junction has not been anyhow developed and got, in fact, totally forgotten.

In this paper, a broad approach is presented, linking a class of hierarchical aggregation algorithms ("minimum distance" algorithms) with a class of objective functions for clustering (the "bi-partial objective functions"). For a broader treatment of the subject of the bi-partial objective functions we refer to Owsiński (2020), while here we concentrate on the derivation of the corresponding hierarchical merger algorithms.

4. A telling example

For purposes of introducing the framework that we shall promote here, let us remind the objective function for clustering, introduced by Marcotorchino and Michaud (1979, 1982), as a part of the mathematical programming formulation of the clustering problem. This formulation was, after a slight modification, as follows:

$$\text{maximise } \sum_{i,j \in I} (y_{ij} s_{ij} + (1-y_{ij})d_{ij}) \qquad (2)$$

where $d_{ij}$ denote the distances between objects in the data set, and $s_{ij}$ – the similarities (proximities) between them, while the decision variables are $y_{ij} = 1$ when objects *i* and *j* belong in the solution to the same cluster, and $y_{ij} = 0$ when they belong in the solution to different clusters, this formulation being subject to the following obvious constraints:

$$y_{ij} \in \{0,1\}, \forall\, i, j,$$

$$y_{ij} = y_{ji}, \forall\, i, j, \text{ meaning symmetry} \qquad (3)$$

$$y_{ij} + y_{jv} - y_{iv} \leq 1, \forall\, i, j, v, \text{ meaning transitivity.}$$

Of course, formula (2) could be written down with an explicit expression, accounting for the transformation $s(d)$ or vice versa, but, both for purposes of preserving generality, and intuitive appeal, as well as simplicity of formulation, we keep here and onwards explicitly both quantities, that is – distance (*d*) and proximity (*s*), assuming, of course, that there is a transformation between them that can be applied in definite concrete cases.

---

[2] We mean here, of course, only the so-called „internal" criteria or indices, the „external" ones making reference to some given partitions, treated as references.

Formulation (2), (3) addresses, definitely, both the adequate representation of the clustering problem, as also formulated here (i.e. the possibly strong internal cohesion of clusters and the possibly distinct separation of clusters), and the way of approaching the solution. The obvious numerical problem arises in connection with the transitivity constraint, which requires $O(n^3)$ inequalities to be treated.

The objective function (2) can, however, be parameterized (see Owsiński and Zadrożny, 1986, 1988) in the manner shown below:

$$\max_P Q(P,r) = r\Sigma_q \Sigma_{i<j \in Aq} s_{ij} + (1-r)\Sigma_{q<q'}\Sigma_{i \in Aq} \Sigma_{j \in Aq'} d_{ij} \tag{4}$$

where we no longer use the decision variables $y_{ij}$, but refer to the partition $P$ as the "decision variable", and where the parameter $r \in [0,1]$ reflects the weight, assigned the two parts of the objective function, now denoted $Q(P,r)$, these two parts corresponding, respectively, to the *internal proximity of objects inside clusters* and to the *distances between objects from various clusters*. It is, naturally, tacitly assumed that the "proper solution" is found for $r = ½$ (i.e. when (4) is equivalent to (2)).

Assume now a procedure, meant to find the optimum solution in terms of partition $P$, based on moving the value of $r$, say, from 0 towards 1. Thus, for $r = 0$, the formulation of the problem, given no other constraints, except for those enforcing the partition, (3), would yield the solution, in which each of the objects constitutes a separate cluster, i.e. $p$ (the number of clusters in $P$) = $n$, since the first component in (4) simply disappears. Given the constraints (3), such a solution is feasible, and shall indeed be provided by any (correct!) method whatsoever.

Now, as the value of the parameter $r$ is increased from 0, the very first encountered obvious solution to (4), satisfying the constraints (3), and, at that, different from the one, obtained for $r = 0$, would appear to be the one, which merges two most similar (least distant) objects $i^*$ and $j^*$ (i.e. such that $d_{i^*j^*}$ = $\min_{i,j} d_{ij}$). The switch from the optimum partition for $r = 0$, which we shall denote $P^*(r=0) = P^*(0) = P^0$, with $P^0 = I$, to the one, in which objects indexed $i^*$ and $j^*$ form a two-object cluster, takes place at a definite value of the parameter $r > 0$. Denote this value, at which the optimum $P^0$ is replaced by $P^1$, established by the merger of $i^*$ and $j^*$, by, accordingly, $r^1$ (implying that $r^0 = 0$).

Let us note that, quite obviously, if there exist in the data set the pairs $(i,j)$, for which $d_{ij} = 0$, then the value of (4) and (2) does not change, whether we merge these objects, or not. Actually, we can assume that for such special cases $r^1 = r^0 = 0$, with $r^1$ corresponding to $P^1$, the partition, which incorporates the merger of all the objects, among which all the distances are zero. An analogous reasoning applies when there are more equidistant objects than just pairs at the level of minimum distance. Hence, from now on, just in order to omit triviality (while, in fact, entirely preserving the validity of the reasoning), we shall be considering that the distances among the objects in the data set are different from 0, and that all distances are different as to their values.

Thus, the (first) merger of a pair $i^*$ and $j^*$ (for which $d_{i^*j^*} = \min_{i,j} d_{ij} > 0$) takes place for the parameter value $r^1$, determined by $d_{i^*j^*}$ and $s_{i^*j^*}$. In order to show the relation, take the two partitions, $P^0$ and $P^1$, the latter one differing from the former by just one merger, and the values of objective function for these partitions, $Q(P^0,r)$ and $Q(P^1,r)$. Hence, we compare the values of

$$Q(P^0,r) = r\Sigma_i s_{ii} + (1-r) \Sigma_{i<j} d_{ij} \tag{5}$$

and

$$Q(P^1,r) = r(\Sigma_i s_{ii} + s_{i^*j^*}) + (1-r)(\Sigma_{i<j} d_{ij} - d_{i^*j^*}) \tag{6}$$

for the parameter $r$ increasing up from 0. Obviously, the values of $\Sigma_i s_{ii}$ and $\Sigma_{i<j} d_{ij}$ are constant for any given set of objects. The comparison (equation) of $Q(P^0,r)$ and $Q(P^1,r)$ yields:

$$r\Sigma_i s_{ii} + (1-r) \Sigma_{i<j} d_{ij} = r(\Sigma_i s_{ii} + s_{i^*j^*}) + (1-r)(\Sigma_{i<j} d_{ij} - d_{i^*j^*}); \tag{7}$$

hence, after simple operations, this leads to

$$rs_{i^*j^*} - (1-r)d_{i^*j^*} = 0 \qquad (8)$$

and we get the sought value of $r^1$:

$$r^1 = \frac{d_{i^*j^*}}{d_{i^*j^*} + s_{i^*j^*}}. \qquad (9)$$

Formula (9) is very telling, indeed. Namely, as we shift the value of the weight parameter $r$ from 0 upwards, we look for the smallest possible value of this parameter, for which $P^0$ is no longer the best partition in terms of (4), and should be replaced by another partition. We conclude from (9) that we obtain this smallest value, $r^1$, exactly for the *smallest distance between two objects*. Formula (9) is valid for any pair of objects, but the value of the merger parameter $r$ is the smallest for the closest two of them.

It is quite straightforward to observe that the reasoning, leading to formula (9) applies to (i) all the subsequent (disjoint) pairs of objects in the set *X*, ranked according to their pairwise distances, and then to (ii) all the subsequent mergers of clusters, no matter how many objects they may contain (in this case the respective formula would account for the appropriate characteristics of individual clusters and of pairs of clusters involved). If, namely, we adopt quite a natural definitions, closely associated with (4), namely:

$$D_{qq'} = \sum_{i \in Aq} \sum_{j \in Aq'} d_{ij} \quad \text{and} \quad S_{qq'} = \sum_{i \in Aq} \sum_{j \in Aq'} s_{ij}, \text{ with } q \neq q', \qquad (10)$$

then the formula (9), for the merger step, indexed $t$, takes the more general form:

$$r^t = \frac{D_{qq'}}{D_{qq'} + S_{qq'}}. \qquad (11)$$

At this point, it should be noted that formulae (9) and (11) provide a simple rule for merging the objects and clusters, analogous to the rules of the classical agglomerative schemes. This particular rule, expressed by (9) and (11), is equivalent to the average linkage, as noted already in Ducimetière (1970).

Yet, here, we do not only proceed as in these agglomerative schemes, guided by a definite merger rule, but dispose also of the "global" objective function, which allows for the evaluation of the entire successive partitions obtained. Obviously, the iterative merger procedure, which is proposed here, is only a suboptimising one, but an improvement is always possible, through certain additional procedures, although, naturally at a cost. We look for the ultimate solution at the value of $r = ½$.

As a complement to the global objective function and its values, we get the sequence of values of $r^t$, constituting a natural index of the hierarchy, with values obtained in the course of the procedure, more closely associated with the objective function than the usually applied distance value.

5. A more general perspective

Having in mind the previously presented example, we can now try to outline a general approach. Thus, quite in analogy to (4) we can propose solving the problem

$$\max_P Q(P,r) = rQ_S(P) + (1-r)Q^D(P) \qquad (12)$$

where $Q_S(P)$ represents a measure of internal similarity of clusters, extended over the entire partition *P*, and $Q^D(P)$ represents a measure of distances between clusters, also extended over the entire partition *P*. It appears quite natural to propose a "dual" to (12) in the form of

$$\min_P Q'(P,r) = rQ^S(P) + (1-r)Q_D(P) \qquad (13)$$

where $Q^S(P)$ reflects the inter-cluster similarity measure, extended over the entire partition, and $Q_D(P)$ reflects the intra-cluster distance measure, also extended over the entire partition (see Owsiński, 2020, for a broader treatment). Yet, we shall continue considering the form (12), the reasoning, concerning the "dual" of (13), being fully analogous.

In order to proceed, we are obliged to make some assumptions, concerning $Q_S(P)$ and $Q^D(P)$. Yet, it is obvious that, in any case, the choice of $Q_S(P)$ and $Q^D(P)$ must be made, first of all, with consideration of the clustering-oriented rationality. Hence, whatever the limitations, introduced by any assumptions, they ought to be assessed from the standpoint of this rationality. And, obviously, this rationality should be assessed with respect to $Q_S(P)$ and $Q^D(P)$ jointly, not separately, as it is often the case in various clustering algorithms.

On the other hand, there are more than one line of reasoning, and hence of the assumptions, leading to our goal, i.e. derivation of the family of minimum distance merger rules, analogous to those of the classical hierarchical merger algorithms, from the objective function like (12). Hence, we shall concentrate here on the examples thereof, in order not to extend the paper too much, and to show that the general form assumed allows for quite a margin of flexibility. In any case, it will be quite apparent that the illustrative case, presented in the preceding section, definitely follows the precepts introduced for the general case.

At the most general level, we shall only assume that the functions $Q_S(P)$ and $Q^D(P)$ are characterised by the *opposite monotonicity along any hierarchy of partitions, arising from mergers*. This is, of course, not the same as opposite monotonicity with respect to $p$, the number of clusters, unless we add to this condition some appropriate qualification, like, $\max_P Q_S(P|\text{card}P = p)$ and, analogously, $\max_P Q^D(P|\text{card}P = p)$.[3]

Hence, for any hierarchy of partitions $H = \{P(p)\}_p$ (any dendrogram, resulting from the merging procedure) we can propose that

$$\arg \min_p Q^D(P \in H) = 1, \text{ and so } \arg \max_p Q^D(P \in H) = n, \tag{14}$$

therefore:

$$\arg \min_p Q_S(P \in H) = n, \text{ and so } \arg \max_p Q_S(P \in H) = 1. \tag{15}$$

The above is insofar intuitively plausible as, indeed, there are no inter-cluster distances for $p = 1$ (first part of (14)), and there are no intra-cluster distances nor proximities for $p = n$ (first part of (15)).

Now, let us return to (12) and put $r = 0$:

$$\max_P Q(P,0) = 0 \cdot Q_S(P) + 1 \cdot Q^D(P) \tag{16}$$

which means, according to (14), that $P^*(0) = I$, where $P^*(r)$ is the best partition, in terms of (12), for the given parameter value $r$. Thus, for $r = 0$ we get the (optimal) solution, in which each object is a separate cluster. Assume we increase $r$ from 0 and try to maximise (12) along the values of this parameter. An alternative to the solution $P^*(0)$, which, definitely, stays valid for $r$ small enough[4] is provided by a partition $P$, different from $I$, for which, conform to the assumption adopted, $Q_S(P) > Q_S(I)$ and, at the same time, $Q^D(P) < Q^D(I)$. Thus, it becomes obvious that for all $P$ different from $I$ a value of $r$ can be found, denoted $r^*(P,I)$, for which $P$ becomes a better partition, in terms of (12), than the partition $I$:

$$r^*(P,I) = \frac{Q^D(I)-Q^D(P)}{Q^D(I)-Q^D(P)+Q_S(P)-Q_S(I)} . \tag{17}$$

---

[3] Other possibly introduced assumptions would concern the nature of the sequences of values of $Q_S(P)$ and $Q^D(P)$ for consecutive mergers (e.g. monotonicity of differences between the consecutive values), and/or, especially, the association with the values of inter-cluster and intra-cluster distance / proximity measures.

Definitely, $r \in [0,1]$. Also, the values of $r^*(P,I)$ are smaller, when the difference $Q_S(P)-Q_S(I)$ is bigger.

Let us remind that in line with the algorithmic scheme, considered in this study, we assume that we look for the subsequent partitions, forming dichotomous hierarchies, i.e. for the sequences $P^t$, $t = 0, 1,...,n-1$, such that $P^t$ is established by the merger of a pair of clusters, forming partition $P^{t-1}$. We can put $P^0 = I$, in accordance with the above, and so (17) becomes

$$r^*(P^1,P^0) = \frac{Q^D(P^0)-Q^D(P^1)}{Q^D(P^0)-Q^D(P^1)+Q_S(P^1)-Q_S(P^0)} . \tag{18}$$

It is obvious that the same relation holds for each subsequent partition along the index $t$:

$$r^*(P^t,P^{t-1}) = \frac{Q^D(P^{t-1})-Q^D(P^t)}{Q^D(P^{t-1})-Q^D(P^t)+Q_S(P^t)-Q_S(P^{t-1})} . \tag{19}$$

Definitely, *relations (17)-(19) hold* not just for the dichotomous hierarchy and for its immediately neighbouring levels (partitions), but *for any two partitions coming from the same hierarchy, with appropriate preservation of the order of the two*.

Here, however, more narrowly, we look, given some $P^{t-1}$, for the "best" among the $P^t$, formed by the merger of two clusters, composing $P^{t-1}$, the "best" being expressed in terms of (12) (ultimately, for $r = \frac{1}{2}$). Resulting from these consecutive choices is the sequence (hierarchy), denoted $\{P^{*t}\}$. We shall propose that it arises from the minimisation, for each consecutive $P^{*t-1}$, of the respective $r^*(P^t,P^{*t-1})$, i.e.

$$P^{*t} = \arg \min r^*(P^t,P^{*t-1}), \tag{20}$$

with $P^t$ being limited to partitions formed by a merger of two clusters, composing $P^{*t-1}$. We shall further denote $\min r^*(P^t,P^{*t-1})$ by $r^t$.

Minimisation from (20), given (19), is equivalent to maximisation of the quotient

$$\frac{Q_S(P^t)-Q_S(P^{*t-1})}{Q^D(P^{*t-1})-Q^D(P^t)}, \tag{21}$$

meaning that the merger ought, in a greedy manner, exploit the biggest current benefit in terms of $\Delta Q_S(P^t) = Q_S(P^t) - Q_S(P^{*t-1})$ and $\Delta Q^D(P^t) = Q^D(P^{*t-1}) - Q_S(P^t)$.[5]

---

[4] As already mentioned, we assume that there are no zero inter-object distances in the set considered, although existence of such zero distances (identical objects) is not a problem for the entire reasoning (all of the respective objects can be aggregated already at $r = 0$).

[5] As we refer to the minimum distance procedure, a natural implication arises, concerning the properties of the bi-partial objective function, as to the relation between $\Delta Q_S(P^t)$ and $\Delta Q^D(P^t)$, on the one hand, and $\min_{q,q'} D(q,q')$, on the other hand.

Let us, then, consider, in this context, the shape of the potentially resulting function $Q(P,r) = rQ_S(P) + (1-r)Q^D(P)$ as the procedure, outlined above, is performed along $r$, starting from $r^0 = 0$. Since the gradient of $Q(P,r)$, with respect to $r$, is $Q_S(P) - Q^D(P)$, we start, for $r^0 = 0$, with $P^0 = I$, from the biggest decrease possible at this step. At the other extreme, for $t = n-1$, we have $P^{n-1} = P^{*n-1} = \{I\}$, meaning all objects in one cluster, and the gradient of $Q(P,r)$ at this end attaining its (positive) maximum. The obvious conjecture is that we deal with a convex, piece-wise linear function, changing its gradient at every merger (every $t$). This is definitely true along the sequence of $P^t$'s, but not necessarily along the values of $r$, as this will depend upon the concrete forms of $Q_S(P)$ and $Q^D(P)$.

At this point we have effectively formulated a general procedure of suboptimisation of the bi-partial clustering criterion, this procedure being very much like the classical hierarchical merger procedures, but with clear indication of the point, when the generation of the dendrogram may be stopped (i.e. $r^t \leq \frac{1}{2}$, $r^{t+1} \geq \frac{1}{2}$). Indeed, the example of this procedure that we have provided in the preceding section was equivalent to one of the classical merger algorithms, which thereby gets equipped with a very definite indicator of partition quality, without the need to recur to "external" statistical measures.

We shall now provide a choice of further examples of such procedures for a couple of selected definitions of $Q(P,r)$.

6. The examples

The precepts that we have adopted for the function $Q(P,r)$ allow for a very broad variety of formulations of this function, and hence also of the potentially associated merger algorithms. We shall now present a couple of examples thereof, starting with the simplest cases, the first one actually bordering upon triviality. In order not to prolong the exposition, we provide only the basic precepts, leaving in most cases the derivation of the concrete form of the key merger relation, analogous to (19), to the Reader.

6.1. The additive objective function with a constant cluster cost

The simple case, presented here (see also Owsiński, 2020) illustrates the possibility of representing the facility location problem in terms of the explicit bi-partial objective function. Notwithstanding the possible variants of the actual facility location problem, we can represent it in the following manner, for the minimised version of the bi-partial function:

$$Q_D(P) = \Sigma_q D(A_q); \quad Q^S(P) = p \tag{22}$$

which, even if a bit artificial from the clustering perspective, regarding especially the form of $Q^S(P)$, if appropriately scaled (normalised), so as to adequately represent the facility location problem, can still be treated as a representation of the bi-partial paradigm. In this case, $Q^S(P)$ represents the total cost of establishing a cluster (a facility location), while $D(A_q)$ represent the costs, associated with distances, generated by the subsets of demand / supply points. The $D(A_q)$ can have, for instance, the form analogous to (10), or the one of the classical k-means algorithm (sum of distances to the cluster centroid).

With this kind of bi-partial function, we can devise a progressive merger procedure, which is based on the smallest distances between clusters. Actually, for $p = n$ we have, assuming that $D(x) = 0$ for any object $x \in E_K$, which is quite natural,

$$Q_D(I) = 0; Q^S(I) = n, \text{ i.e. } Q(I) = n. \tag{23}$$

Then, the question is: can we find a pair of objects, say $i^*$ and $j^*$, such that

$$D(\{i^*,j^*\}) < 1,$$

so that the value of the overall objective function for the partition, in which only this particular pair is formed, is smaller than in (23), i.e. than for the partition $P = I$:

$$D(\{i^*, j^*\}) + n - 1 < n. \tag{24}$$

Consequently, it can be concluded that, as long as the value of $D(A_q)$ for a cluster $A_q$, which is formed through aggregation of any (two) clusters from some preceding partition, is smaller than 1, it pays to proceed with the thus defined aggregation, since $Q(P)$ shall thereby decrease.

In this manner we can easily design in details a progressive merger algorithm that would stop once no longer $D(A_q) < 1$ can be found, meaning that we established a suboptimal solution in terms of $Q(P)$. It suffices for the function $D(A_q)$ to satisfy quite natural and simple conditions (no decrease of the value being possible as we join the objects and clusters that are increasingly more distant) for the thus outlined procedure to determine a good approximation of the optimum solution.

The above not only justifies the incorporation of this case in the bi-partial paradigm, but also indicates the essential issue of relevance for the facility location problem, namely the issue of appropriate scaling of values, so that the whole problem, and hence its solution, have a practical (or at least interpretative) sense.

6.2. The case of minimum distances and maximum proximities

Another reasonable representation of the clustering problem might involve the following, actually quite classical, definitions of distances and proximities for the particular clusters:

$$D(A,B) = \min_{i \in A, j \in B} d_{ij}; \text{ and } S(A) = \max_{i,j \in A} s_{ij} \tag{25}$$

with the following definitions of the components of bi-partial objective function:

$$Q^D(P) = \Sigma_q \Sigma_{q'>q} D(A_q, A_{q'}) \text{ and } Q_S(P) = \Sigma_q \text{card}A_q \cdot S(A_q), \tag{26}$$

where $\text{card}A$ denotes the number of objects in the set $A$.

This formulation, while offering a relatively well justified model for the clustering problem, allows also for the use of the progressive merger algorithm, very similar to those already proposed, and also leading to a suboptimal solution. The formulation satisfies by itself the conditions for such an algorithm to be applicable, and to stop at the suboptimal solution.

6.3. The case of average distances and additive proximities

In this case it is possible to apply the progressive merger algorithm, based on the direct application of the precepts analogous to those already presented. Here, at the level of individual clusters, we refer to the following definitions:

$$D(A,B) = \frac{1}{\text{card}A \cdot \text{card}B} \sum_{\substack{i \in A \\ j \in B}} d_{ij}, \text{ and } S(A) = \frac{1}{2} \sum_{i,j \in A} s_{ij}, \tag{27}$$

while the respective two components of the overall bi-partial objective function, i.e. $Q^D(P)$ and $Q_S(P)$, are simple summations over all clusters, forming the partition.

6.4. The k-means case: the objective function

We shall now present the bi-partial version of the classical k-means algorithm and its setting. The version here introduced differs from the original one by the formulation of the objective function, with the consequences, concerning the nature of solutions obtained, and, most importantly, the possibility of specifying the number of clusters, this being routinely done for the classical k-means exclusively using the "external" criteria.

Let us remind that the general formulation of the classical k-means is based on the minimised objective function that we shall denote in the usual manner here adopted, $Q_D(P)$. In the standard setting of the classical k-means algorithm, $Q_D(P)$ has the form

$$Q_D(P) = \Sigma_q \Sigma_{i \in A_q} d(x_i, x^q), \qquad (28)$$

where $d(.,.)$ is some distance function[6], $x_i$ is a vector of values, characterising object $i$, and $x^q$ is the (analogous) characterisation of the "representative" of cluster $A_q$, $q = 1,...,p$.

The minimum values of $Q_D(P)$ are obtained with the use of the primeval generic algorithm associated with the k-means, namely the "centre-and-reallocate" one:

for some initial, possibly random, set of $x^q$, $q = 1,...,p$, assign the elements of the set $X$ to the closest $x^q$, determining thereby the clusters $A_q$, then calculate the new $x^q$, e.g. as averages over $A_q$, and go back to the assignment step, stopping the entire procedure whenever there is no change between the iterations, or the change satisfies definite conditions.

It is known that this algorithm very efficiently leads to a local minimum of $Q_D(P)$ for a given value of $p$ – in a very limited number of iterations for even quite large data sets.

In addition, even though it is known that the "centre-and-reallocate" procedure attains just a local minimum of the given objective function, starting the procedure from multiple randomly generated sets of "representatives" $x^q$ ultimately yields the proper minimum for the given number of clusters $p$. There exist a number of techniques for improving the choice of initial set of $x^q$, which diminish the number of necessary repetitions of the procedure.

The minimum values of $Q_D(P)$, determined with the primeval algorithm for consecutive numbers of clusters, $p$, these values denoted here as $Q^*_D(p)$, decrease, from a simile of total variance, conform to (28), for $p = 1$, i.e. for the entire set of objects considered, treated as a single cluster, down to zero for $p = n$ (or even "earlier", in terms of $p$). Hence, it is necessary to run the respective algorithm, with repetitions, mentioned before, for several consecutive values of $p$ and apply an external (statistical) criterion, say, AIC, BIC, Calinski & Harabasz, etc., in order to pick the "proper" number of clusters $p$.

In this manner we are obliged, in order to find the "best" $p$, to recur to a criterion that does not stem from the same "philosophy" as the objective function we used, $Q_D(P)$, and the respective algorithm, and thus a criterion applied may be poorly fit to the original k-means framework (the criterion may be oriented at a different character of partition than produced by the k-means algorithm). This important reservation holds true even if we accept the results produced by such an "external" criterion on the basis of comparison of results produced by several different criteria (all of them being actually in principle similarly "external" to the original procedure).

We would like to resolve this issue by applying the bi-partial approach. The weak point of this approach is that there is no ready recipe for designing the concrete forms of $Q^S_D(P)$. It is quite often so that – like in the case of k-means type of algorithms – some objective function, whether explicit or implicit, corresponding to an existing approach, and representing either $Q^S(P)$ or $Q_D(P)$ (i.e. $Q^D(P)$ or $Q_S(P)$ for the dual formulation), can be complemented with an appropriate counterpart, which has then to be cleverly designed.

---

[6] Actually, in order to ensure the fulfilment of the appropriate properties of the k-means algorithm, especially related to the securing that the cluster mean minimises the sum of distances to the objects in a cluster, mostly the squared Euclidean distance is referred to in this context. Otherwise, the basic properties mentioned are replaced by respective approximations.

The reward, however, may be worth the effort: for the pair of functions fulfilling certain additional conditions, which in some cases are quite natural, one obtains also a straightforward, even if not always very effective, algorithm, leading to the optimum or sub-optimum partition $P$.

In this particular case, the objective function, matching the principles of k-means, might have the following minimised form:

$$Q_D^S(P) = Q_D(P) + Q^S(P), \quad (29)$$

with $Q_D(P)$ defined as in formula (28). The second component, $Q^S(P)$, would then have to reflect the overall measure of inter-cluster proximity (similarity). It might take, in particular, the form of

$$Q^S(P) = \tfrac{1}{2} \sum_q S^*(A_q), \quad (30)$$

with $S^*(A_q)$ being the "outer similarity" measure for the cluster $A_q$, defined, say, as

$$S^*(A_q) = \sum_{i \in A_q} \max_{q' \neq q} s(x_i, x^{q'}). \quad (31)$$

With the definitions, provided through the formulae (29)-(31) we can observe the behaviour of the objective function upon choosing a definite measure of distance. If we choose Manhattan, or city-block distance, we deal in the "centering" step only with an approximation of the minimum sum of distances. Computations, carried out for simple academic data sets, serve to verify the basic features of the proposed objective function. An example thereof is shown in Fig. 1. It can be easily seen that it is meant to check the possibility of indicating more than one "best" partition, and the respective (alternative / potential) partitions are well visible. It can be said that in this example one deals with true "nested" partitions (two or even three levels of quasi-optimum partitions).

This example, on the one hand, provides the possibility of checking the "correctness" of results from a clustering algorithm, and, on the other, poses challenges that cannot be properly addressed by the existing clustering methods.

The values of the functions, involved in (29), obtained in respective consecutive solutions for the successive values of $p$, are provided in Table 2.

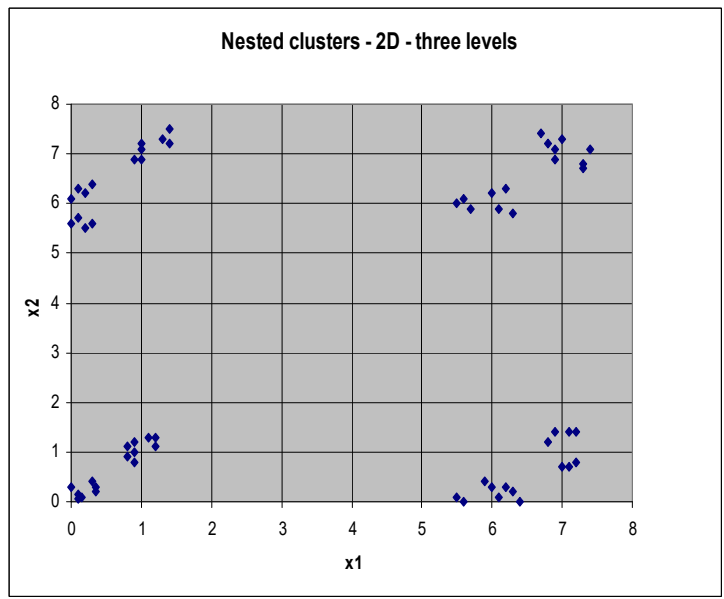

Figure 1 An academic example of a data set for the bi-partial k-means

Some explanations are due, concerning the exemplary calculations. Thus, $n = 60$. The values of the objective function $Q_D^S$ for the extreme numbers of clusters (1 and 60), shown in Table 2, differ, despite the application of an "averaging" definition of transformation $s(d)$ (i.e. transformation, which preserves the average value over the set of pairs of objects), due to the rounding, appearing in the actual transformation. Finally, the values for $p = 7$ were not calculated.

**Table 2.** Optimum values of the objective function (29) and its components for the successive numbers of clusters for the academic example, illustrated in Fig. 1

| No. of clusters | $Q_D^S$ | $Q_D$ | $Q^S$ |
|---|---|---|---|
| 1 | 353.65 | 353.65 | 0 |
| 2 | 338.87 | 207.05 | 131.82 |
| 3 | 333.67 | 134.87 | 198.80 |
| 4 | 265.15 | 59.67 | 205.48 |
| 5 | 288.23 | 51.05 | 237.18 |
| 6 | 307.38 | 42.52 | 264.86 |
| 7 | X | X | X |
| 8 | 335.40 | 21.96 | 324.44 |
| 60 | 361.23 | 0 | 361.23 |

Concerning the results obtained, let us note that the bi-partial objective function, as designed for this particular case, indicates "correct" values of $p$ as corresponding to the partitions (clusterings), visible in the data sets. So, a very clear minimum is observed for $p = 4$. No proper local minima, though, are observed, as it could be hoped, for $p = 8$, and possibly higher values of $p$. Still, for $p = 8$ a slight deviation is observed in the respective curve of $Q_D^S(p)$ from the upward trend towards $Q_D^S(60)$.

These other minima would have been obtained if a different transformation $s(d)$ were applied (see also the next subsection) or a different weight were used in formula (30). The assumptions made appear, however, to be quite "natural", and these other solutions can be reasonably sought only after the initial, "natural" solution would have been found.

6.5. The k-means case: the procedure

We shall now show the development of the corresponding algorithmic procedure for the k-means case. Let us remind that we deal here with the objective function

$$Q_D^S(P) = Q_D(P) + Q^S(P) \text{ with } Q_D(P) = \Sigma_q \Sigma_{i \in Aq} d(x_i, x^q) \text{ and } Q^S(P) = \tfrac{1}{2} \Sigma_q \Sigma_{i \in Aq} \max_{q' \neq q} s(x_i, x^{q'}), \quad (32)$$

which is minimized.

Now, as before, we parameterize this function:

$$Q_D^S(P,r) = rQ_D(P) + (1-r)Q^S(P) \quad (33)$$

and start with, say, $r^0 = 1$. For this value of $r$ we have $\min_P Q_D^S(P,1) = \min_P Q_D(P)$, and thus it is obvious, given the properties of $Q_D(P)$, that the optimum $P(r^0) = P^0 = I$, as $Q_D(I) = 0$.

Starting with $r^0 = 1$ we decrease the value of the parameter, and so $(1-r)Q^S(P)$ increases from zero (while $rQ_D(P^0)$ decreases). The part of the objective function, associated with $Q^S(P)$, is equal

$$(1-r) \tfrac{1}{2} \Sigma_{i \in I} \max_{j \neq i} s_{ij}. \quad (34)$$

Conform to the idea adopted, we look for the possibly highest (first encountered) value of $r$, for which $P^0$ is replaced by another partition, minimizing the objective function. We look for the shift, constituted by the merger of two objects (clusters), but, in fact, at this stage (parameter value) this would be a global minimum. So, we select for the merger the objects (clusters) $i^*, j^*$, on the basis of $\max_{i,j} s_{ij}$ (i.e., $\min_{i,j} d_{ij}$), and compare the values of the objective function before and after the merger, which takes place, when the left hand side (before) is bigger than the right hand side (after):

$(1-r) \frac{1}{2} \Sigma_{i \in I} \max_{j \neq i} s_{ij} > r \min_{i,j} d_{ij} + (1-r) \frac{1}{2} (\Sigma_{i \in I} \max_{j \neq i} s_{ij} - s_{i^*j^*})$. (35)

After straightforward transformations, we get from (35):

$r < s_{i^*j^*}/(s_{i^*j^*} + d_{i^*j^*})$, (36)

which, definitely, confirms our choice of $s_{i^*j^*} = \max_{i,j} s_{ij}$, since we look for the first, i.e. the biggest $r$, satisfying (36). We have thus obtained the relation, corresponding to (17), actually analogous to (9).

Now, we would like to ascertain that when we deal with clusters composed of more objects, the relations, derived before, also generally hold.

So, again we compare the values of the objective function before and after merger:

before: $r \Sigma_q D(A_q) + (1-r) \frac{1}{2} \Sigma_q S(A_q)$ (37a)

after: $r (\Sigma_q D(A_q) - D(A_{q1}) - D(A_{q2}) + D(A_{q12})) + (1-r) \frac{1}{2} (\Sigma_q S(A_q) - S(A_{q1}) - S(A_{q2}) + S(A_{q12}))$,

(37b)

where clusters indexed $q1$ and $q2$ are supposed to be merged and to form together the cluster indexed $q12$. Like before, when looking for the condition that (37a) gets bigger than (37b), after simple transformations, we obtain:

$$r < \frac{\frac{1}{2}(S(A_{q1})+S(A_{q2})-S(A_{q12}))}{\frac{1}{2}(S(A_{q1})+S(A_{q2})-S(A_{q12}))+(D(A_{q12})-D(A_{q1})-D(A_{q2}))}.$$ (38)

Again, we try to find the biggest $r$ by selecting appropriate clusters, indexed $q1$, $q2$. So, we have formulated a progressive merger algorithm, which fits into the k-means framework, but actually proceeds in a different manner.

At the end of this section we shall yet devote some attention to the properties of the quantities, which appear in formula (38), namely the crucial two:

$\Delta S_{q1q2} = S(A_{q1}) + S(A_{q2}) - S(A_{q12})$ (39a)

$\Delta D_{q1q2} = D(A_{q12}) - D(A_{q1}) - D(A_{q2})$. (39b)

First, it is obvious that both of these are non-negative for any two clusters, indexed $q1$ and $q2$. Hence, the value of (38) belongs to the interval [0,1]. Then, also in connection with the previous statement, (39a) can be regarded as a measure of proximity between two clusters, while (39b) – as a measure of distance between two clusters. Hence, we definitely deal here with the minimum distance aggregation rule.

Another interesting point is the resulting sequence of values of $r^t$ for subsequent cluster mergers. Although we choose the possibly biggest value, according to (38), we cannot be assured that the sequence we get is monotonic (non-increasing). Yet, if it so occurs for some merger $t$ that the value of $r^t$ is bigger than $r^{t-1}$, then we have a clear indication that the distribution of objects in space is inhomogeneous in the manner, illustrated in Fig. 1 (groups of close-by clusters, these groups being separated by larger distances).

6.6. Some computational issues

We have presented here the examples of both the formulations of the bi-partial objective function and the associated merger algorithms. As mentioned already, the hierarchical merger algorithms suffer from the computational burden, which makes them of little or no use for really large data sets. That is why nowadays, such approaches are used, if at all, either for smaller data sets, for "didactic" purposes, or in conjunction with other, faster approaches, in the form of hybrid procedures. This is often done with a faster algorithm forming initial groups, which are then merged by a hierarchical merger algorithm. This is, for instance, done in Owsiński and Mejza (2007) and Owsiński (2010), with classical k-means in the first stage, which starts from a relatively high number of clusters, say $n^{1/2}$, with the bi-partial objective values $Q^S_D(P)$ being observed for consecutive, decreasing $p$. Then, the mergers are performed according to the bi-partial-generated rules.

The case of k-means, described before, is insofar specific as by applying the bi-partial-based merging we lose, of course, the advantage of the original k-means in terms of fast convergence, without the need of taking into account all the distances / proximities.

Yet, altogether, the verification of applicability of the bi-partial approach to the k-means-type paradigm shows that it is not only fully feasible, but also effective. Definitely, similar (or analogous) bi-partial functions can be defined also for the fuzzy version of the paradigm (FCM). The paper by Dvoenko and Owsiński (2019) presented yet another version of the same paradigm ("the meanless k-means"), also transformed using the bi-partial approach.

7. Conclusions

The paper introduced the general outline for a broad class of progressive merger procedures of clustering, based on the concept of bi-partial objective function, which serves to represent correctly (fully) the original problem of clustering. It is shown how the bi-partial objective function can be designed for a variety of circumstances (assumptions as to the clustering criteria), and how the merger algorithms can be derived for these formulations, approximating the corresponding optimal solution. Thus, the approach provides, first, the effective instrument of evaluation of the partitions / clusterings obtained, and also an effective, general form of respective suboptimisation procedure.

In view of the computational issues, related to the merger algorithms, it is advised to recur to hybrid algorithms, with mergers performed in the later stages of such algorithms. It must be noted, though, that the broad applicability and versatility of the overall approach allows for very different forms of the merger rules, some of which may be much more computationally efficient than the others. Another open question relates to the choice and consequences of the more strict conditions on the clustering criteria, which could be used within the bi-partial framework regarding the correct construction of the objective function and the derivation of the merger algorithms.

Compliance with Ethical Standards

Funding: no special funding is associated with this paper

Conflict of Interest: there are no conflicts of interest, associated with this paper and its contents

Ethical Conduct: this paper does not involve any relevant aspects from the point of view of ethical conduct

Data Availability Statement: this paper refers solely to small academic sets of illustrative data, which can be obtained from the author upon request